# A survey on proximity monitoring and warning in construction

Yuexiong Ding[a], Qiong Liu[b], Ankang Ji[a], Xiaowei Luo[b, *], Wen Yi[a, *], Albert P.C. Chan[a]

[a]Department of Building and Real Estate, The Hong Kong Polytechnic University, Hong Kong, China

[b]Department of Architecture and Civil Engineering, City University of Hong Kong, Hong Kong, China

[*] Corresponding author: Xiaowei Luo (xiaowluo@cityu.edu.hk) and Wen Yi (wen.yi@polyu.edu.hk)

## Abstract

Various technologies have been applied to monitor the proximity between two construction entities, preventing struck-by accidents and thereby enhancing onsite safety. This study comprehensively reviews related efforts dedicated to proximity monitoring and warning (PMW) based on 97 relevant articles published between 2010 and 2024. The bibliometric analysis reveals the technical roadmap over time, as well as the five most influential leaders and the two largest research networks they have established. The qualitative review is then conducted from four perspectives: influencing factor study, hazard level definition and determination, proximity perception, and alarm issuing and receiving. Finally, the limitations and challenges of current proximity perception are discussed, along with corresponding future research directions, including end-to-end three-dimensional (3D) object detection, real-time 3D reconstruction and updating for dynamic construction scenes, and multimodal fusion. This review presents the current research status, limitations, and future directions of PMW, guiding the future development of PMW systems.

## 1. Introduction

The construction industry is widely regarded as one of the most dangerous industries worldwide. "Struck-by" accidents, a leading cause of injuries and fatalities in the construction industry [1], occur when a moving object, piece of equipment, or material hits a worker. According to statistics, approximately 75% of fatal collisions are related to construction equipment [2]. Keeping a safe distance from construction entities (e.g., workers, equipment, etc.) can help reduce the occurrence of related accidents and casualties. Therefore, real-time monitoring of the proximity between any two construction entities for early collision warning is an effective way to reduce struck-by accidents, which has always been one of the key issues that need to be addressed in construction safety management [3].

Over the past few decades, various methods have been proposed to monitor or estimate proximity, thereby preventing collisions between construction entities and improving the safety of construction sites. Traditional methods rely on manual monitoring, which is too labor-intensive and costly to be conducted continuously and is susceptible to human subjectivity. The application of Internet of Things (IoT) and wireless communication technologies enables real-time and continuous monitoring, such as radio frequency identification (RFID) [4,5], ultra-wideband (UWB) [6,7], Bluetooth [8,9], global positioning system (GPS) [10,11], etc. In recent years, with the rapid development of computer vision (CV) and deep learning, CV-based methods have been widely applied for collision prevention monitoring [3,12–14], as surveillance cameras offer the advantages of a wide coverage range, non-object-specificity, and a non-contact nature.

Though many efforts have been dedicated to developing proximity monitoring and warning (PMW) systems based on various emerging technologies, comprehensive reviews of this topic are still lacking. Nonetheless, there are still some similar review studies [15,16]. For example, Ye et al. [16] reviewed the performance evaluation of struck-by-accident alert systems in road construction areas. Demeke et al. [15] reviewed the application of sensor technologies for intrusion and proximity hazards monitoring in highway work zones. These existing related reviews are limited to the field of road construction, which is only a part of the construction industry. Meanwhile, sensor technology is a part of the technologies used for proximity warning. Moreover, these reviews lack a comprehensive investigation from the perspectives of key processes involved in PMW system development.

Therefore, this study aims to systematically review various efforts committed to the development of PMW systems. A mixed-methods review combining quantitative and qualitative analysis is adopted to provide complementary insights from different perspectives. Specifically, a bibliometric analysis is first conducted to identify the technical focuses over time and the influential groups and authors in the field. Followed by a qualitative review, covering four primary aspects of the PMW system, including influencing factor study, hazard level definition and determination, proximity perception, and alarm issuing and receiving. Finally, the limitations and challenges of current proximity perception are discussed, along with corresponding future research directions.

The contributions and significance of this study can be summarized as follows: 1) A comprehensive review of PMW is conducted from both quantitative and qualitative perspectives; 2) The bibliometric analysis uncovers the technological evolution roadmap of PMW over time, as well as the most influential groups and authors in the field; 3) The qualitative review provides a detailed classification and summary of the methods, technologies, strategies, or modes involved in the key processes of PMW system development, providing readers with a general framework for PMW system development; 4) Finally, several promising research directions for the future are proposed based on the discussion regarding limitations and challenges of current proximity perception, which are expected to guide the further development of PMW systems in the future.

The rest of the paper is organized as follows: Section 2 outlines the research methodology. Section 3 presents research focuses and trends identified through bibliometric analysis. Section 4 provides a qualitative review of the collected articles from four aspects of developing proximity warning systems. Section 5 highlights the limitations and challenges in current research, as well as potential future research directions. Finally, Section 6 concludes the study.

## 2. Methodology

A widely used mixed-methods review [17,18] that combines bibliometric analysis and qualitative review is adopted in this study to explore the current research status of PMW in construction. Figure 1 shows an overview of the research framework, including 1) publication retrieving and collection from two indexing platforms, 2) bibliometric analysis to form a scientific map of existing works, and 3) qualitative review from four perspectives to gain a comprehensive understanding of current efforts.

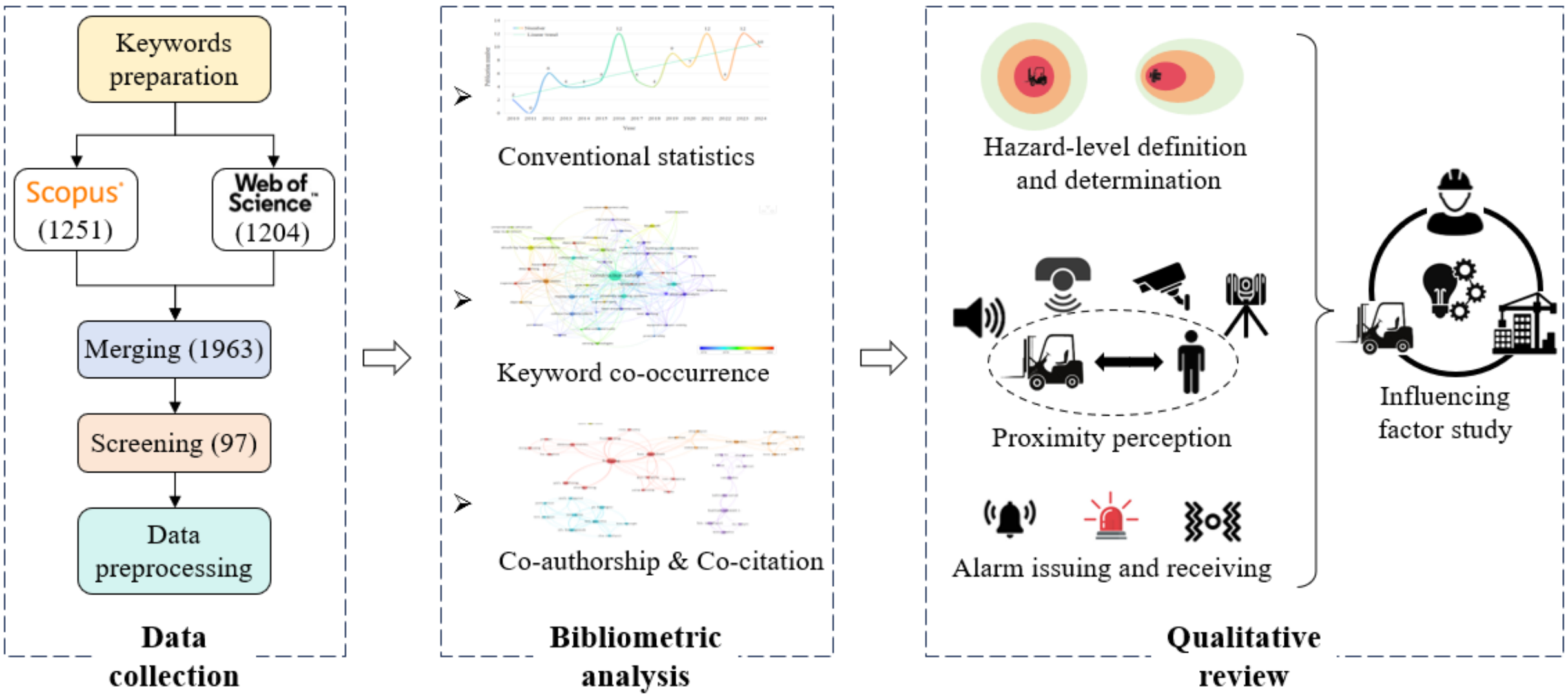

Figure 1. Research framework.

### 2.1 Data collection

Two famous databases, Web of Science (WoS) and Scopus, were utilized to retrieve relevant publications using the advanced Boolean search combination of construction-related keywords (CRK) and target topic-related keywords (TRK), as shown in Table 1. The search field was "Title/Abstract/Keyword", and the duration was from 2010-01-01 to 2024-12-31.

Other restrictions were set as follows: *Document Type = "Article", Language = "English", Subject Area = "Engineering"*. Please refer to Table A1 for the complete list of search strings.

After retrieval, 1,251 and 1,204 articles were obtained from Scopus and WoS, respectively. The collected articles were further reduced to 1963 after merging and removing duplicates. Finally, 97 articles with strong relevance to the target topic were selected by manual screening, using two criteria: 1) the research addressed proximity hazard-related issues, and 2) the application field was construction-related. Some data preprocessing operations were performed to improve the quality of input data of bibliometric analysis, including the unification of data format (the WoS format), author keywords (refer to Table A2), and author names (e.g., "Heng Li", "Li, Heng", and "H. Li" refer to the same author).

Table 1. Two types of keywords for literature retrieval.

| | |
|---|---|
| Construction-related keywords (CRK) | worker*, architecture, engineering, construction, aec, infrastructure, site, building, civil |
| Topic-related keywords (TRK) | str?ck, str?ck-by, "collision detection", "collision monitoring", "collision prediction", "collision warning*", "collision alert*", "proximity detection", "proximity monitoring", "proximity prediction", "proximity estimation", "proximity warning*", "proximity alert*", "proximity perception", "proximity awareness", "distance detection", "distance monitoring", "distance prediction", "distance estimation", "distance perception", "distance awareness", "distance prediction", "proximity hazard*", "proximity sensing", "proximity analysis", "proximity alarm*", "collision hazard*", "collision avoidance", "collision accident*", "collision alarm*", "intrusion warning*", "intrusion alarm*", "intrusion alert*", anticollision, anti-collision, "collision risk*", "dangerous proximity", "hazardous proximity", "unsafe proximity" |

## 2.2 Bibliometric analysis

Bibliometric analysis is a scientific computer-aided review method that visualizes the internal relationships of publications related to a specific topic, presenting the overall research trend and focus of the field. VOSviewer [19] is a widely used tool for generating and visualizing bibliometric networks, and its powerful capabilities in bibliometric analysis have been demonstrated in numerous previous review studies [17,18]. Therefore, this study utilized VOSviewer to quantitatively analyze the keyword co-occurrence, co-authorship, and co-citation, thereby revealing the scientific map of existing works.

### 2.3 Qualitative review

A qualitative review provides a deeper understanding of existing research. As shown in the right part of Figure 1, the qualitative review encompasses the entire process of PMW system development, from the preparation stage to the proximity perception stage and then to the alarm issuing stage, providing a comprehensive overview and in-depth discussion of the selected publications. The insights gained during this process can help identify the limitations of current research, inspire advancements in the field, and provide potential solutions to address existing gaps.

## 3. Bibliometric analysis

### 3.1 Overview

Figure 2 shows the annual number of publications from 2010 to 2024. Note that the number of publications in 2024 may be much higher than the number counted in this study, as some of the relevant publications originally scheduled for 2024 may not have been included in WOS and Scopus at the time of the last retrieval in this study.

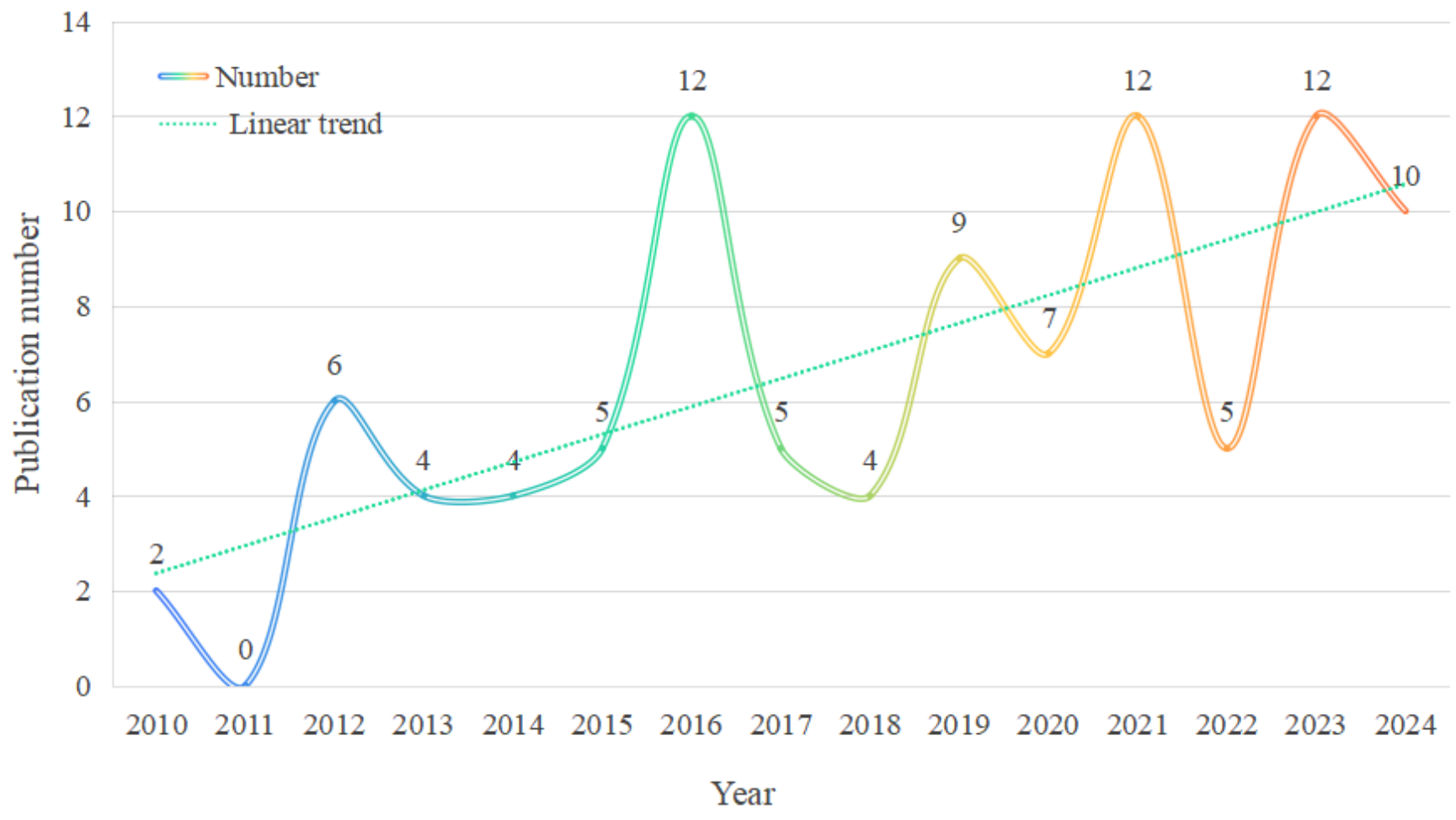

Figure 2. Number of publications over the year.

Overall, despite fluctuations, the annual publication number shows an upward trend. Specifically, the curve gradually increases from 2010, reaches its peak for the first time in 2016, and then begins to decline sharply. Starting from 2018, the curve exhibits a fluctuating upward trend, reaching its peak in 2021 and again in 2023. This phenomenon typically indicates a transition from old to new methods/technologies, i.e., the old reach a bottleneck and the new start gaining attention and becoming prevalent. The transition will be further reflected and discussed in the keyword co-occurrence analysis in Section 3.2. Additionally, there is a notable decline between 2021 and 2023. Such short-term violent fluctuations are more likely to be caused by global emergencies. For example, the isolation, remote work, and shutdown strategies adopted in response to the COVID-19 pandemic during 2020-2021 may lead to delays in related research projects, ultimately resulting in a significant decline in publications.

Table 2 provides statistics on the source publication and the corresponding citations. The top three journals, ranked by the number of articles, are *Automation in Construction*, *Journal of Computing in Civil Engineering*, and *Journal of Construction Engineering and Management*. These three journals account for 54% of the total publication volume, demonstrating their leading position in the field. In addition to the number of publications, the average citation can also reflect the journal's influence, where the top three journals are *Automation in Construction*, *Journal of Information Technology in Construction*, and *Journal of Computing in Civil Engineering*.

Table 2. Journal publications and citations.

| **Source** | **Article number** | **Citations** | **Average citations** |
|---|---|---|---|
| Automation in Construction | 22 | 2220 | 100.9 |
| Journal of Computing in Civil Engineering | 16 | 878 | 54.9 |
| Journal of Construction Engineering and Management | 14 | 387 | 27.6 |
| Sensors | 7 | 136 | 19.4 |
| Safety Science | 6 | 179 | 29.8 |
| Advanced Engineering Informatics | 5 | 251 | 50.2 |
| Computer-Aided Civil and Infrastructure Engineering | 3 | 145 | 48.3 |
| Applied Sciences | 3 | 66 | 22.0 |
| Buildings | 3 | 10 | 3.3 |
| Journal of Information Technology in Construction | 2 | 116 | 58.0 |
| Sustainability | 2 | 60 | 30.0 |
| Frontiers in Built Environment | 1 | 27 | 27.0 |
| Accident Analysis and Prevention | 1 | 19 | 19.0 |
| Journal of Sensors | 1 | 15 | 15.0 |
| Proceedings of the Institution of Mechanical Engineers, Part D: Journal of Automobile Engineering | 1 | 13 | 13.0 |
| IEEE Transactions on industrial Electronics | 1 | 11 | 11.0 |
| Journal of Mechanical Science and Technology | 1 | 10 | 10.0 |
| Transportation Research Record | 1 | 8 | 8.0 |
| Electronics | 1 | 4 | 4.0 |
| Journal of Industrial Information Integration | 1 | 3 | 3.0 |
| Transportation Research Part C: Emerging Technologies | 1 | 3 | 3.0 |
| Journal of Robotics and Mechatronics | 1 | 1 | 1.0 |
| Sensors and Transducers | 1 | 1 | 1.0 |
| Measurement: Journal of the International Measurement Confederation | 1 | 0 | 0.0 |
| Mechatronic Systems and Control | 1 | 0 | 0.0 |

### 3.2 Keyword co-occurrence

Keyword co-occurrence analysis refers to revealing research focus and trends in the field by analyzing the frequency of keywords appearing in the literature. High-frequency keywords usually represent research hotspots, while the co-occurrence frequency of keywords reflects the correlation between different research topics.

The co-occurrence network overlaid with the keyword occurrence year was generated to explore research focus in different periods, as shown in Figure 3. Combined with the annual publication curve in Figure 2, the evolution of the technological roadmap for proximity monitoring and warning over time can be uncovered:

- Stage I (2010-2016, in purple). This stage mainly utilized "information technologies" such as "RFID", "laser scanning", and "point cloud" for relevant "real-time" analysis. Though "laser scanning" and "point cloud" technologies are currently the most popular and advanced technologies for 3D spatial perception, they have been rarely applied in this field since 2016. The main reason is that it is challenging to balance high performance and low cost to meet the requirement of high computational power for these technologies to achieve "real-time" analysis. Therefore, these technologies have not been extensively explored; instead, they have been applied as emerging methods for non-real-time operations, such as data acquisition and preprocessing [20,21].

- Stage II (2017-2019, blue to green). At this stage, advanced sensor technologies with higher precision, such as UWB and Bluetooth low energy (BLE), are gradually replacing RFID for positioning. Additionally, new technologies such as unmanned aerial vehicles (UAVs), deep neural networks, virtual reality (VR), and building information modeling (BIM) were also gradually being tentatively applied during this stage.

- Stage III (2020-present, yellow to orange-red). Transitioning through Stage II, the dominance of "deep learning" and "computer vision" -based technologies, such as "object detection", "object tracking", "trajectory prediction", etc., has gradually been established and is now reaching its climax. In addition, a new "vibrotactile warning" method has also attracted much attention during this stage.

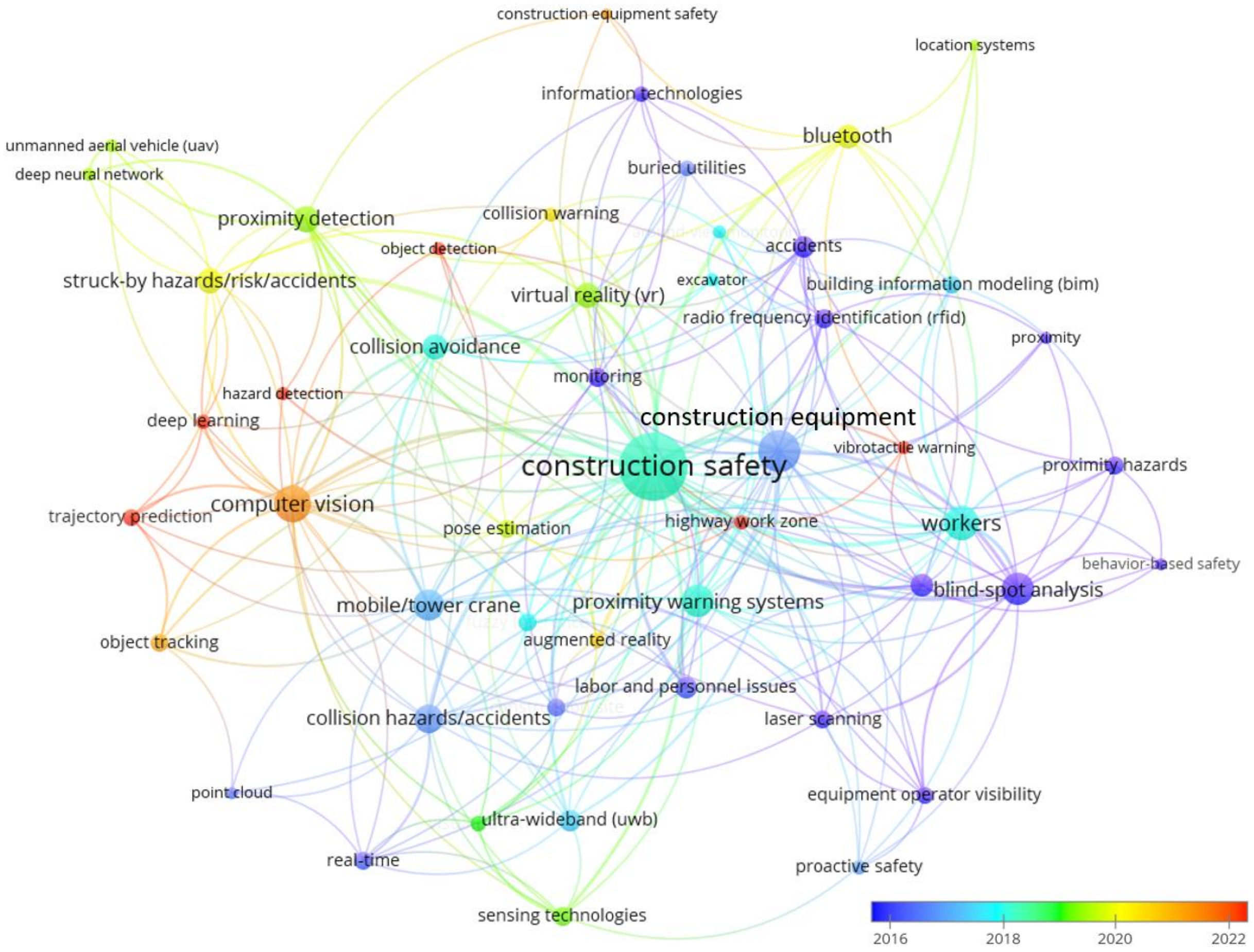

Figure 3. Keyword co-occurrence network overlaid with the keyword's average occurrence year.

### 3.3 Co-authorship and co-citation

Co-authorship and co-citation analysis can both be used to identify active and influential authors, organisations, and countries within a specific field. Figure 4 displays the core part of the co-authorship network, focusing on the top 6 sub-networks with the most nodes. Each node represents an author, and the connections between nodes indicate their collaboration. The node size represents the number of publications. There are two significant collaborative sub-networks in Figure 4. The research teams led by Jochen Teizer (8) and Yong K. Cho (6) formed the largest sub-network (29 members), followed closely by the sub-network centered around Heng Li (23 members). As for individual output, prolific authors include Jochen Teizer (8 publications), Heng Li (7), Yong K. Cho (6), Amin Hammad (6), Vineet R. Kamat (4), Xiaochun Luo (4), Eric D. Marks (4), and Tao Cheng (4).

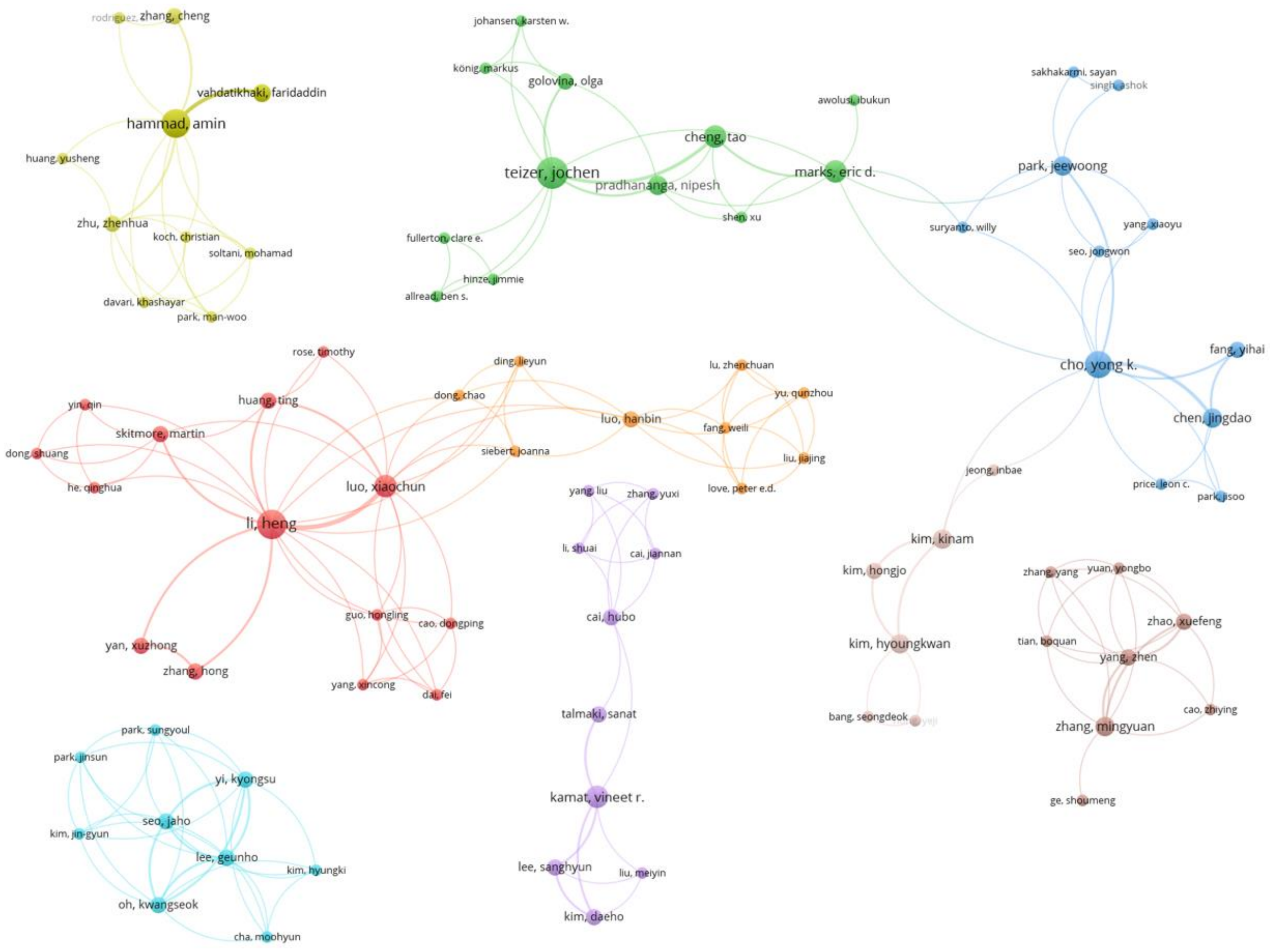

Figure 4. Co-authorship network.

Co-citation refers to papers that are cited together in third papers, indicating a similar research focus between the co-cited papers. Based on author co-citation analysis, the top 10 co-cited authors, sorted by total link strength, are listed in Table 3. Five influential authors overlap with those authors in Figure 4, including Jochen Teizer, Heng Li, Yong K. Cho, Xiaochun Luo, and Tao Cheng. This overlap reveals their significant academic contributions and influence in the target field.

Table 3. Top 10 co-cited authors.

| **Author** | **Co-citations** | **Total Link Strength** |
|---|---|---|
| Teizer, Jochen | 366 | 12486 |
| Kim, Hyoungkwan | 131 | 5202 |
| Li, Heng | 123 | 5016 |
| Cho, Yong K. | 95 | 3400 |
| Hinze, Jimmie | 83 | 3284 |
| Lee, Sanghyun | 83 | 3107 |
| Caldas, Carlos H. | 61 | 2993 |
| Cheng, Tao | 61 | 2961 |
| Luo, Xiaochun | 65 | 2828 |
| Luo, Hanbin | 59 | 2796 |

## 4. Qualitative review

As shown in the right part of Figure 1, this section categorizes and summarizes the collected publications according to four key procedures for the system development of PMW: 1) influencing factor study, 2) hazard-level definition and determination, 3) proximity perception, and 4) alarm issuing and receiving.

### 4.1 Influencing factor study

The investigation of influencing factors aims to identify potential factors that can reduce or increase the likelihood of human-machine collision accidents. Fully considering these influencing factors will help develop more comprehensive and efficient warning systems. After categorization, as shown in Figure 5, these influencing factors mainly include human-related, equipment-related, and environmental factors.

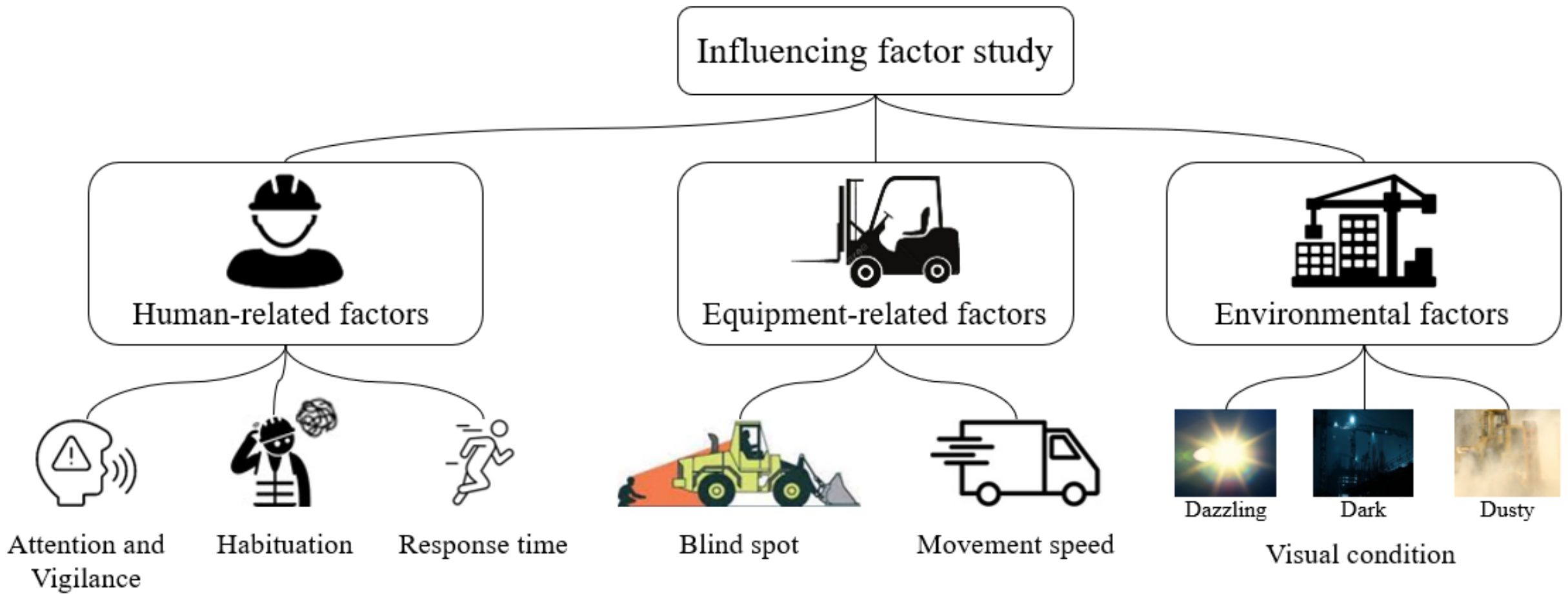

Figure 5. Commonly influencing factors mentioned in the selected articles.

#### 4.1.1 Human-related factors

Human-related factors mainly refer to workers' attention to surroundings, awareness of danger (vigilance), habituation to alarms, and response time to alarms. Ideally, workers should always be aware of and alert to nearby and approaching hazards. However, they are usually too focused on their work to keep sufficient attention and vigilance towards their surroundings. Attracting workers' attention or predicting their inattention behaviors in the early stage can effectively reduce struck-by hazards caused by inattention. To this end, Eiris et al. [22] trained workers using three different online media (static images, cinemagraphs, and videos) to explore the effectiveness of these media in guiding workers' attention for recognizing struck-by hazards. Kim et al. [23] found significant differences in biological signals between workers' attention-focused and inattentive behaviors. Then they developed machine learning methods to predict workers' inattentive behaviors based on workers' biological signals. Hussain et al. [24] utilized eye tracking technology to explore the relationship between workers' attention and awareness in various virtual collision scenarios, aiming to help safety trainers understand the changes in workers' attention and vigilance over time, thereby reducing the risk of accidents caused by inattention.

Long-term exposure to high-risk environments or repeated high-risk operations can lead to the development of risk habituation, which is another crucial factor contributing to a decline in attention and vigilance. Understanding the development of risk habituation and providing effective interventions is crucial for preventing accidents from occurring. Kim et al. [25] investigated the effectiveness of VR technology as a behavioral intervention tool in mitigating the decline of workers' vigilance resulting from risk habituation. They found that simulated accidents can continuously mitigate the effects of risk habituation on workers' attention over a one-week interval. Chae et al. [26] quantified habituation to auditory warnings using workers' behavioral and physiological data. Based on this quantification method, they [2] then explored the effectiveness of three different alarm sounds of construction equipment in reducing workers' habituation to auditory warnings. Additionally, receiving too many redundant or inaccurate alerts may also lead to workers developing a risk habituation. Therefore, Chan et al. [27] incorporated workers' field of vision as a proxy for their risk awareness into the generation of hazard proximity warnings to reduce the occurrence of redundant or inaccurate alarms.

Workers' reactions and response times to alarms vary. Therefore, studying workers' reactions and response time can make alarm systems more scenario- or job-specific. Luo et al. [28] proposed a novel method to evaluate workers' responses to proximity warnings of static safety hazards, finding that carpenters experienced longer response delays in dangerous areas than steelworkers. Based on prior knowledge of workers' responses, they [29] further generated estimated response rates (ERRs) as primary indicators to validate the effectiveness of the given control measures and previously identified proximity hazards. In addition, alarm issuing and receiving methods are also crucial factors that have a particular effect on workers' reactions and response times, which will be further discussed in Section 4.4

#### 4.1.2 Equipment-related factors

Research on equipment-related factors primarily focuses on the equipment's blind spot and movement speed. Due to the unique structure of the equipment cab, the operator's view of specific areas around the equipment is limited. These areas with limited visibility are known as blind spots of the equipment. Blind spots can lead to various accidents, including collisions, crushing, and overturning. To this end, Teizer et al. [1] utilized a 3D laser scanner to measure and identify blind spots of common heavy equipment, such as trucks, excavators, and graders. Subsequently, Marks et al. [30] proposed and validated a blind spot measurement method based on laser scanning data, which not only evaluates and compares different equipment's design features but also demonstrates how the design of equipment affects operator visibility. In addition to common mobile equipment, Cheng and Teizer [20] applied laser scanning technology to the blind spot recognition of tower cranes.

The equipment's movement speed is one of the potential factors that affect the accuracy of distance perception, as too fast speeds may cause inconsistent performance, signal transmission delays, and even malfunction of the detection device. Martin et al. [31] designed a series of experiments to explore the main factors affecting the distance measurement of ultrasonic ranging sensors, including equipment speed, distance between detectors and equipment, and the angle between sensor beams and equipment trajectories. Park et al. [32] proposed parameter adjustment and adaptive signal processing (ASP) methods to mitigate the inconsistent alarm distance and time delays of Bluetooth sensors resulting from high movement speeds. Thapa and Mishra [33] applied survival analysis to identify the key determinants of work zone crashes in the presence of alarm systems, including vehicle speed, distance between sensors and workers, and the accuracy of intrusion detection and alerts.

#### 4.1.3 Environmental factors

The surrounding environment of the workplace can also impact the incidence of accidents, including blind spots created by obstacles, low visibility due to adverse weather conditions, and glare and dimness caused by inappropriate lighting. Adverse construction environments are often accompanied by high accident rates [34], as poor visual environments make it more difficult for workers to notice hazards. Hong et al. [35] conducted a screen-based virtual experiment using eye tracking technology to investigate the impact of visual environments (e.g., normal, dazzling, dark, and dusty conditions) on device operators' recognition of impact hazards. They found that the poor visual environment significantly weakened operators' hazard detection ability, especially in low illuminance and low luminance contrast consitions.

### 4.2 Hazard level definition and determination

As shown in Figure 6, hazard level definition and determination include two steps: 1) customize various hazard levels for construction entities (workers or equipment) based on specific scenarios, and then 2) determine the hazard levels of entities based on their status and locations during construction.

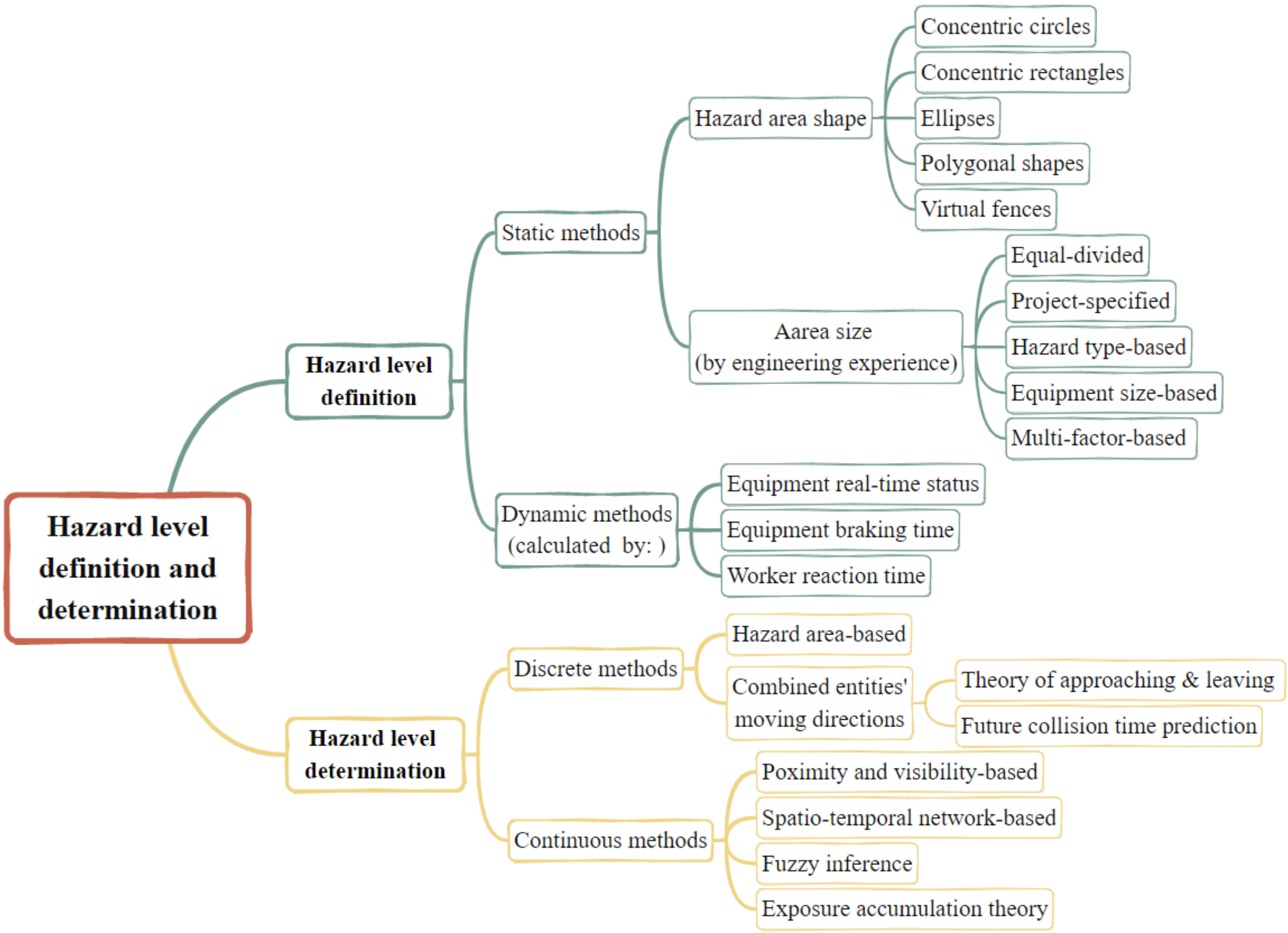

Figure 6. Mind map of hazard level definition and determination.

#### 4.2.1 Hazard level definition

*1) Static methods*

Static methods refer to maintaining consistency in the shape and size of the area designed for the hazard level throughout the entire construction activity. As shown in Figure 7, these area shapes can be concentric circles/fans [1,21,36,9,37–41,7], concentric rectangles [13], ellipses [42], polygonal shapes [43], and polygonal virtual fences [44–46]. The concentric circle scheme has been adopted by the vast majority of studies, which simply treats objects within the same circular area equally without considering their specific locations. However, this scheme overlooks many factors that affect the area design, such as the equipment's attributes (e.g., size, speed, direction, turning radius, braking distance, etc.) and the worker's field of view. To this end, Gan et al. [13] designed the concentric rectangle scheme, which takes into account the rectangular dimensions of certain construction vehicles. Considering more equipment attributes, Shen et al. [43] customized irregular polygonal shapes for specific heavy construction vehicles. These polygonal shapes are static because the attribute values are pre-estimated and set for a specific equipment. Additionally, the worker's field of view is also valuable for area shape design. Shin and Kim [42] found that the risk of accidents among workers is lower in their binocular area (front side), higher in their monocular area (left and right sides), and most significant in the blind spot (back side). Based on this, they designed the ellipse scheme that embeds multiple elliptical shapes, with small front area and large back area. Finally the virtual fence refers to manually demarcating polygonal regions in images [45,46] or visual interfaces [44] for monitoring construction entities' entry and exit in these regions via techniques such as object detection or sensor positioning.

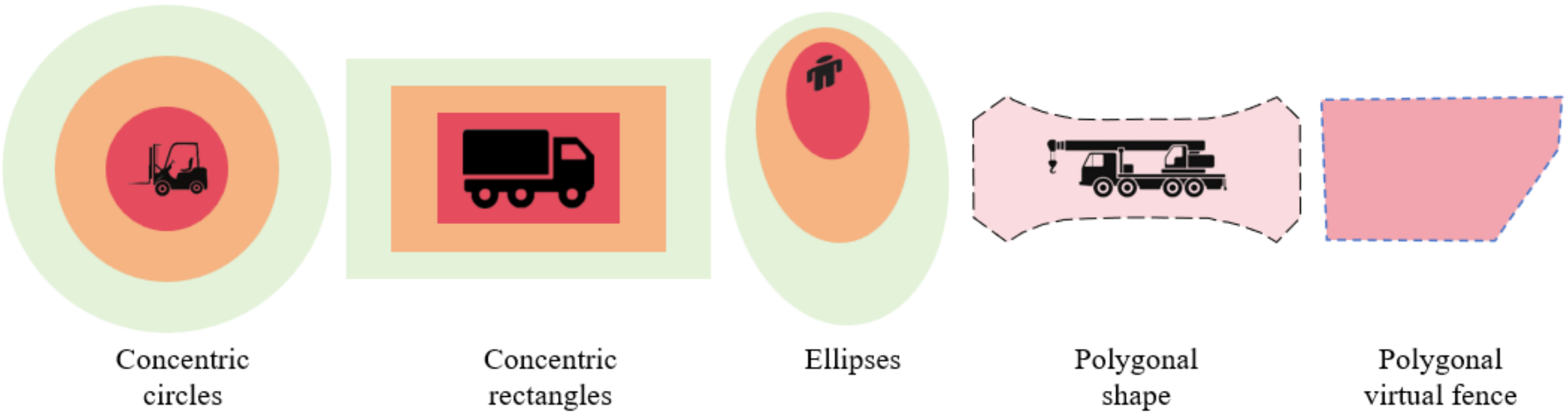

Figure 7. Examples of various types of static area shapes.

The sizes of these static area shapes are mainly determined based on the engineering experience of researchers or managers, which can be categorized into equal-divided [36,39,40], project-specified [44–46], hazard type-based [21], equipment size-based [7,13,37,38,41], and multi-factor-based [42,43] methods. For example, Sakhakami et al. [39] used concentric circles with radii of 4, 8, and 12 meters to represent high-risk, medium-risk, and low-risk zones. Virtual fences [44–46] must be designed according to the specific content and requirements of the construction project. Different hazards may have different distance requirements for proximity. For example, according to industry best practices, Fang et al. [21] assigned distance thresholds of 4, 2, and 1 meters for power lines, building/equipment, and trees, respectively. As for heavy construction vehicles, they are usually extended outward for a certain distance based on their size and operating radius. For example, Son et al. [37] set distance thresholds of 2, 6, and 12 meters for their excavator to identify hazard levels from high to low. In addition to equipment size, blind spots [42], equipment turning radius and braking distance, and the operator's reaction time [43] can all affect the setting of these area sizes.

*2) Dynamic methods*

Static methods ignore real-time status changes of construction entities, resulting in a large amount of redundant space reserved for each targeted object, thereby reducing their applicability in crowded construction scenarios. Instead, dynamic methods take these real-time statuses into account to update hazard areas promptly [14,47–50], i.e., the area shapes and sizes change dynamically. For example, Vahdatikhaki and Hammad [47,48] identified the dynamic workspaces of earthwork equipment as danger areas, which were updated in real-time based on the equipment's current status (e.g., posture, speed) and the required braking time. Tian et al. [50] defined the hazard areas of an excavator as spaces that all its active components will sweep over in a future time interval (e.g., the average reaction time of the driver). Seong et al. [14] generated a dynamic egg-shaped area for mobile equipment based on the equipment's size and speed, as well as workers' speed and their reaction time.

### 4.2.2 Hazard level determination

*1) Discrete methods*

The hazard level can be represented as a discrete category (i.e., discrete methods) or a normalized continuous value (i.e., continuous methods). Discrete methods consider construction entities in the same area as having the same hazard level. For example, consider all entities within the red zone as having the same highest danger level, as shown in Figure 7. Most studies in Section 4.2.1 employed this method because it is pretty straightforward. However, this method ignores entities' moving directions, i.e., whether entities are approaching or moving away from each other, resulting in a large number of false alarms. Therefore, Wang and Razavi [51,52] developed a new strategy that triggers alarms only when the distance between entities is less than the threohold and they are getting closer to each other. Similarly, Huang et al. [53] classified workers within the same area as having different hazard levels based on their moving directions. Furthermore, the future trajectory and collision time can also be predicted based on the moving direction, allowing for the early determination of hazard levels for a certain time interval in the future. For example, Oh et al. [54,55] identified three hazard levels using the time-to-collision (TTC) and warning index, where the TTC is predicted based on the movement status of entities and the warning index is defined by the distance between entities.

*2) Continuous methods*

Continuous methods utilize continuous numerical values to represent the hazard degrees of entities. These numerical values can be calculated based on the entities' proximity, physical attributes, and real-time status (e.g., speed, location, posture, etc.). For example, Vahdatikhaki and Hammad [47] proposed a novel method based on proximity and visibility (i.e., the time that the area is not within the equipment's blind spot) to generate risk heat maps by integrating equipment status with near-real-time simulation. Wang and Razavi [56] quantified the interactions between entities according to their proximity, blind spots, and speeds to generate a spatio-temporal network, which can be used to calculatie the risk degree of a single entity (node/entity-level analysis) or the entire construction site (network-level analysis). Fuzzy inference can also be used for calculating risk degrees. For example, Kim et al. [57] and Zhang et al. [58] first fuzzified the proximity and crowding of workers using predefined membership functions, then inferred the results using predefined if-then rules, and finally used a defuzzification to convert the inference results into continuous risk degrees.

Some studies assume that the hazard degree of an entity can be calculated based on its frequency [49] and duration [11,59] of exposure to hazardous areas. For example, Teizer and Cheng [49] proposed the proximity hazard indicator to represent the degree of risk, which indicates how often the observed entity is exposed to predefined hazards within a fixed observation period. Based on the gridded spatiotemporal GPS data, Golovina et al. [11] first calculated the total score for each grid according to multiple predefined parameters and weights. Then they divided the total score by the total observation time to obtain the corresponding hazard index for each entity. Luo et al. [59] regarded the hazard as a radiation source. They believed that working and moving in hazardous areas is a process of accumulating radiation (i.e., risk), and the likelihood of accidents occurring increases over time. Based on this theory, they proposed the concept of hazard exposure, which can be calculated by combining real-time location data of construction entities with specific hazards of different radiation types (e.g., point, line, and surface radiation).

### 4.3 Proximity perception

Regardless of the method used for hazard level determination, the distance between two entities must be estimated first, known as proximity perception. Therefore, the essential task of a collision warning system is to perceive the proximity of entities in real-time. As shown in Figure 8, the proximity perception methods utilized in the selected articles mainly include sound-based, sensor-based, vision-based, point cloud-based, and multimodal-based methods.

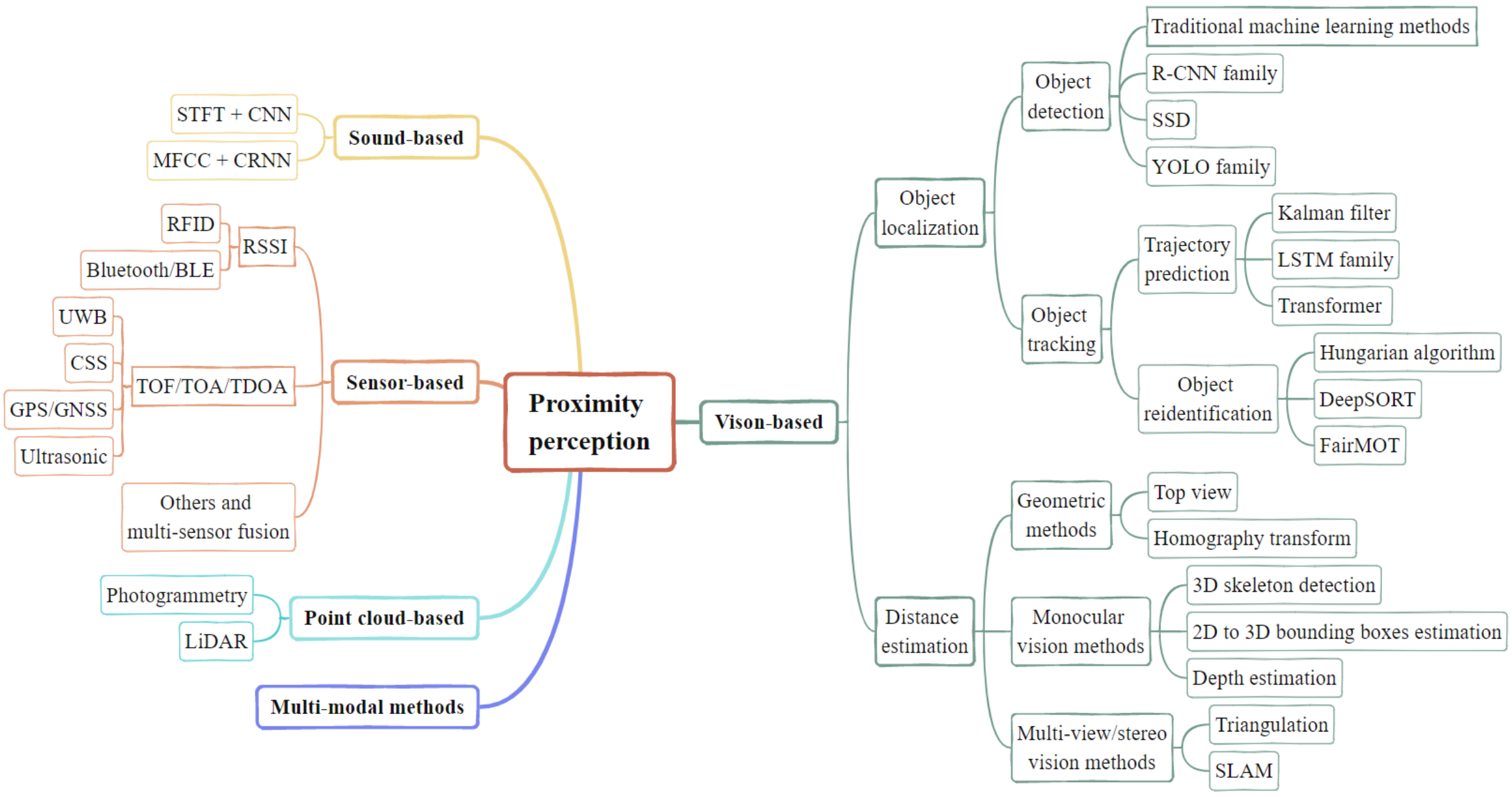

Figure 8. Mind map of proximity perception methods.

### 4.3.1 Sound-based methods

The running construction vehicles typically emit loud and recognizable roaring sounds. According to the Doppler effect, the pitch of the roaring sound increases as the vehicle approaches the observer, and decreases as it departs from the observer. Based on this principle, Lee and Yang [60] converted the obtained acoustic information into a spectrogram through bandpass filtering and a short-time Fourier transform (STFT), and then used a convolutional neural network (CNN) to identify the equipment's movement state, direction, and speed. After applying Mel frequency cepstral coefficients (MFCC) for acoustic features extraction, Elelu et al. [61] also utilized a CNN to identify whether heavy vehicles are moving within a certain distance range in noisy construction environments. They [62] further trained a two-stage convolutional-recurrent neural network (CRNN) model for event sound classification and localization, where the first stage learns multi-label sound events and the second stage estimates their direction of arrival.

In addition to human-perceivable sounds, ultrasound is more effective at detecting distances. Unlike the above methods of monitoring equipment's roaring sound, ultrasound-based methods typically require the equipment to be equipped with an ultrasound generator and receiver. Therefore, ultrasound-based methods will be classified as sensor-based methods.

### 4.3.2 Sensor-based methods

*1) Radio frequency identification (RFID)*

RFID is an automatic identity recognition technology. A typical RFID system consists of tags, readers, and servers. A tag, known as a responder, is a microchip storing unique identity information. A reader is used to query and capture identity information sent by tags, while a server is typically a computer that receives and forwards identity information. According to the power supply, RFID tags are typically categorized into passive [4,5], semi-passive, and active [1,63]. Distances between tags and readers can be estimated by analyzing the signal response or the received signal strength indicator (RSSI). For example, Chae and Yoshida [63] utilized active RFID to demarcate working spaces, thereby restricting the entry and exit of specific entities to prevent human-vehicle collisions. Teizer et al. [1] estimated proximity based on the signal strength of active RFID with very high frequency (VHF). Afterwards, they [4] developed a novel battery-free sensing and communication prototype, a SmartHat with self-monitoring alert and reporting technology for hazard avoidance and training.

*2) Bluetooth/Bluetooth low energy (BLE)*

Bluetooth is a wireless communication technology that enables the exchange of data over short distances. Bluetooth technology enables real-time connections to multiple devices through self-organizing networks, allowing for bidirectional communication between various platforms. BLE has several advantages over traditional Bluetooth, including low cost, low energy consumption, and minimal infrastructure requirements. BLE has sub-meter level accuracy [32] that estimates the distance between transmitter and receiver [8,9,32,36,41,53,64,65] mainly by measuring the RSSI. The effective detection range of BLE is independent of the angle between the transmitter and receiver, but increases with the increase of transmission power [65]. BLE allows any device equipped with the Bluetooth transmission protocol to communicate with it. For example, the proximity warning system developed by Baek and Choi [9] receives signals from BLE beacons attached to miners or vehicles through smartphones for distance estimation. To mitigate the adverse impact of the noise, an effective signal-processing method is necessary. For example, Park et al. [32] developed an adaptive signal-processing (ASP) method, while Kim et al. [41] employed particle filtering for noise reduction in RSSI to enhance the distance estimation performance of BLE. BLE can also achieve global positioning by installing multiple BLE beacons within the target region. For example, Arslan et al. [8] installed approximately 200 BLE beacons in two buildings to detect intrusion behavior by analyzing the spatiotemporal trajectories of workers in the constantly changing buildings.

*3) Ultra-wideband (UWB)*

UWB is also a radio technology that transmits data across a wide bandwidth, commonly used for precise positioning and tracking, with a high precision of around 20 centimeters [66]. In addition to being more accurate and faster than RFID, UWB is more reliable in metal and crowded environments. A UWB positioning system typically consists of three components: a central hub, base stations (also known as anchors), and tags. UWB positioning can be realized using various algorithms, such as two-way ranging (TWR) based on time of flight (TOF), time of arrival (TOA), time difference of arrival (TDOA), etc. According to the number of anchors, the UWB system can generate two-dimensional (with three anchors) or three-dimensional (with at least four anchors) coordinates for each tag. In practical applications, Hwang [66], Zhang and Hammad [67], and Zhang et al. [68] attached multiple UWB tags with identification numbers to key components of cranes, such as the boom, hook, and loads, for real-time positioning and tracking of multiple cranes to detect potential collisions early. Besides, Teizer and Cheng [49] and Jo et al. [69] mounted UWB tags on helmets to enable real-time positioning of workers onsite. Ansaripour et al. [6] developed and tested a vehicle pose estimation system that utilized four UWB tags placed on the vehicle's body to estimate its location, speed, and orientation. To address the limitation of fixed anchors, Ventura et al. [7] abandoned the absolute positioning of construction entities, instead attaching tags to equipment and anchors to workers. In this case, only the distance between any tag and each anchor needs to be calculated. As tags are attached to the equipment, it constantly knows the distance between itself and any detected worker mounted a anchor.

*4) Chirp spread spectrum (CSS)*

CSS is a spread-spectrum technique that utilizes wideband linear frequency-modulated chirp pulses to encode information, enabling the reliable transmission of data over long ranges. Unlike UWB, which is best suited for precise applications, CSS excels at providing accurate long-range outdoor positioning with a precision between one and two meters. For example, Lee et al. [70], Li et al. [44,71], and Luo et al. [59] utilized CSS technology for the real-time positioning of multiple workers to issue early warnings to those approaching danger. Experiments demonstrated that the CSS-based real-time positioning system has an average precision of 0.8 meters and a response speed of 200 milliseconds [71]. Applying CSS positioning, Liu et al. [72] developed a motion model to predict workers' next locations based on their movement patterns and randomness, thereby estimating the risk of collision.

*5) Global navigation satellite system (GNSS)/Global positioning system (GPS)*

GPS is one of the subsystems of GNSS, a space-based radio navigation and positioning system that provides users with all-weather 3D coordinates, velocity, and time information at any location on the Earth's surface or near Earth space. It does not require additional sensing infrastructure (e.g., base stations) to operate, and only needs to be installed on construction entities to obtain their real-time information. GPS is generally used for outdoor positioning, with a precision greater than 1 meter and higher deviations in crowded environments [73]. For example, Pradhananga and Teizer [73] and Golovina et al. [11] utilized GPS to collect spatiotemporal data of construction moving entities, analyzing their speeds and proximity to estimate collision risk. To improve the precision of traditional GPS, Theiss et al. [10] combined real-time kinematic (RTK) technology to calibrate the GPS data of approaching vehicles, achieving a positioning precision within 0.4 inches.

6) Ultrasonic

Like radio waves, ultrasonic waves emitted from the sensor will be bounced back to the sensor when encountering obstacles. Following this principle, the ultrasonic sensor measures the distance from itself to the obstacle based on the TOF taken for signal transmission and reception. For example, Martin et al. [31] utilized ultrasonic sensors to detect vehicle intrusions around road construction areas, and Yong et al. [74] employed ultrasonic sensors for real-time ranging of obstacles and workers within a 20-meter range. Ju et al. [74] proposed a system architecture for long-distance ultrasonic ranging based on a vibration model, enabling real-time, accurate, stable, and efficient detection of obstacles within a range of 33 meters. However, ultrasound is easily absorbed or deviated due to factors such as the irregular surface of the obstacle or its angle with the sensor, resulting in unreliable distance perception results.

*7) Others*

Other types of sensors, such as magnetic sensors [75,76], laser sensors [77,78], joint/rotation sensors (or encoders) [79–83], and inertial measurement units (IMUs) [21,51,52,84], can also be used for distance perception. For example, Kim et al. [75] installed magnetic sensors on the excavator bucket to estimate the distance between the bucket and underground metal pipelines in real time, achieving an overall error of less than 4.38 millimeters. Lee et al. [78] employed laser sensors to measure the height of tower crane loads. Ren and Wu [83] utilized various joint sensors, including rotation angle sensors, elevation angle sensors, boom length sensors, and windlass position encoders, to monitor and predict the motion and position of the crane in real time based on a weighted linear regression model.

Multi-sensor fusion can compensate for the problems of scenario limitations and large estimation errors caused by using a single type of sensor. For example, Wu et al. [85] developed a system that uses RFID for real-time identity confirmation and ultrasonic sensors for ranging the identified entities. Fang et al. [21] used multiple rotation encoders and IMUs for real-time position tracking of cranes and their loads. Based on real geographic information system (GIS) data of underground assets, Talmaki et al. [80,81] combined GPS and various joint encoders to track and simulate the position and pose of the excavator, thereby preventing damage to underground assets during excavation. Similarly, Liu et al. [84] combined GPS, IMUs, and digital twins for 3D excavation simulation to monitor the relative distance between the excavator and underground pipelines. Wang and Razavi [51,52] used a GPS-aided inertial navigation system that integrated an IMU to achieve positioning precision of less than 0.7 meters. Choe et al. [86] designed a series of tests to evaluate the performance of various sensing technologies for backover accident prevention, including sensor installation, static testing, dynamic testing, and dirty sensor testing. Golovina et al. [87] presented valuable suggestions on sensor fusion, data collection and inference, and personalized feedback through two field cases.

### 4.3.3 Vison-based methods

Visual-based methods typically involve two processes: 1) identifying and locating objects in the image, and then 2) estimating the distance between entities based on given assumptions or visual features of the image.

*1) Object localization*

Object localization mainly involves two computer vision tasks: object detection and object tracking.

*a) Object detection*

Traditional object detection is primarily based on the combination of feature engineering and machine learning methods. For example, Kim et al. first used a Gaussian mixture model (GMM) to separate foreground (i.e., moving objects) from background, and then conducted object detection through morphological operations [57] or support vector machine (SVM) classifiers based on histogram of oriented gradient (HOG) features [88]. Price et al. [79] combined multiple algorithms, including edge detection, the Hough transform, and mean-shift clustering, for object detection.

The application of deep neural networks (e.g., CNNs) eliminates tedious image preprocessing and feature engineering in traditional object detection. Deep learning-based object detection can be either bounding box-based object detection or pixel-level object detection (i.e., instance segmentation) [89–91]. 2D object detection can be divided into two-stage and one-stage methods. Representative models of two-stage methods include region-based CNN (R-CNN), faster R-CNN [37,92,58,3,93], mask R-CNN [90,91], and some of their improved variants [89]. Two-stage methods have high accuracy but relatively low inference speed. One-stage methods directly predict results from the extracted features, thereby improving inference speed. Typical one-stage methods include the single-shot detector (SSD) [93] and the You Only Look Once (YOLO) family, such as YOLOv2 [38], YOLOv3 [42,94,95], YOLOv4 [96,97], YOLOv5 [12–14], YOLOv7 [98], and YOLOv8[99]. For 3D object detection, only a small amount of research has been conducted [3,89,100] under the studied topic. For example, Yan et al. [3] indirectly estimated 3D bounding boxes of the truck via the detected 2D bounding boxes. Shen et al. [89] obtained the 3D bounding boxes of objects based on their point cloud instances, which were segmented from the pseudo LiDAR point cloud according to the detected 2D bounding boxes.

*b) Object tracking*

Based on the initial positions obtained from object detection, object tracking aims to predict the future positions of objects and perform object re-identification (ReID). The Kalman filter is a traditional and classic algorithm for predicting future positions [12,57,88,101], which operates through a cycle of "predict-update". Unlike the Kalman filter, which can only predict single-step trajectories, deep learning-based trajectory prediction enables the prediction of trajectories for a period of time in the future [45,95,102,91,40,13]. For example, Cai et al. [102] proposed a context-enhanced long short-term memory (LSTM) model that considers social behaviors and workplace context information to predict a series of future positions of entities based on historical observations. Tang et al. [45] combined LSTM with a mixture density network (MDN) and introduced a multihead module to predict locations at different future times. Zhang and Ge [40] and Gan et al. [13] utilized the Transformer model to achieve higher accuracy in trajectory prediction compared to other deep-learning models, such as social GAN [95] and social LSTM [102]. To eliminate the steps of context information extraction regarding workplaces and entities' movements, Bang et al. [91] utilized a convolutional LSTM (ConvLSTM) to directly predict the future trajectories of entities, using the segmented binary image sequence as input.

Traditional ReID algorithms, such as the Hungarian algorithm [12], often encounter the frequent identity switching problem when an object remains occluded for an extended period. Deep learning based ReID methods have effectively improved this problem [40,98,13]. For example, Gan et al. [13] and Zhang et al. [98] used a deep simple online and real-time tracking (DeepSORT) model for real-time tracking of construction entities, which uses a simple CNN to extract visual features bounded by the detected boxes and conducts the ReID based on the similarity of the feature vectors (e.g., cosine similarity). Zhang and Ge [40] applied a novel model, fairness multiple object tracking (FairMOT), which combines object detection with ReID to perform end-to-end detection and tracking of workers and crane loads.

*2) Distance estimation*

Distance estimation methods mainly include geometric, monocular vision-based, and stereo vision-based (or multi-view-based) methods.

Geometric methods refer to the application of inverse perspective transformation (also known as Homography) to convert the captured source image into a top view (also known as a bird's-eye view), thereby simplifying distance calculations from 3D space to a 2D plane. After the transformation, the actual distances can be calculated based on the ratio of pixel distances to actual distances (i.e., the scale factor). In addition to applying Homography, the camera can also be installed at heights to directly obtain top or near-top views, such as tower crane heights [57,90] and drone platforms [91,13]. The scale factor used for distance estimation can be predetermined based on the camera installation position [57,90,58], or calculated from a reference object of known size [94,95,38,91,40,13], or calculated from four known points (two or more non-collinear) corresponding to the image and the real world [37,97,14,12,98]. Note that after the transformation, the elevation information will be lost, resulting in significant distance estimation errors when entities are at different altitudes. To this end, Zhang et al. [98]

segmented the monitoring area into multiple sub-regions and conducted the same proximity perception and risk assessment for entities within the same sub-region.

Monocular vision-based methods predict the 3D positions of entities based only on existing monocular images. For example, Yan et al. [92] performed 3D skeleton detection of workers based on monocular images, which were then used as view-invariant features to train a deep neural network (DNN) for distance estimation. Similarly, Tian et al. [50] developed a two-stage Transformer-based method for 3D pose estimation of excavators, aiming to locate the positions and states of various components. Yan et al. [3] indirectly estimated 3D bounding boxes of entities using multiple types of detected 2D bounding boxes, including the top, side, and front/back bounding boxes. Shen et al. [89] utilized a DNN for monocular depth estimation and then converted the depth map into a pseudo LiDAR point cloud for point cloud-based 3D bounding box generation. By introducing two adaptive error correction terms, Yao et al. [99] proposed an adaptive monocular depth estimation model to mitigate the distortion of object pixel size resulting from changes in depth and shooting angle.

Multi-view-based methods refer to technologies that calculate depth or distance based on triangulation from binocular or multi-view images. For example, Zhu et al. [101] applied triangulation to predict the future 3D positions of entities based on their 2D positions detected from two cameras. Jeelani et al. [46] employed simultaneous localization and mapping (SLAM) technology to determine the 3D positions of the workers in the generated map based on the cameras installed on their safety helmets. Wang et al. [96] obtained 3D spatial information of the workplace from binocular images using a channel attention-based lightweight stereo matching model. They then estimated the distance between construction equipment and power equipment by mapping the centroid of the 2D detection box to the obtained 3D space.

### 4.3.4 Point cloud-based methods

Point cloud is one of the mainstream methods for 3D spatial representation. The collected articles primarily discuss two methods for obtaining the point cloud of the built environment or construction entities: laser scanning [20,21,54,79,100] and photogrammetry[79]. Identifying the representative points or 3D bounding boxes of point cloud clusters (or instances) of construction entities can achieve entity positioning and distance measurement. For example, Cheng and Teizer [20] used laser point cloud data to identify and measure blind spots that restrict the visibility of crane operators. Fang et al. [21] and Price et al. [79] used laser scanners to capture the point clouds of the lifting operation site, constructing the built environment around the crane. However, capturing and processing the point cloud is quite time-consuming. For example, Price et al. [79] spent 1.5 hours scanning eight different locations at a single construction site, followed by 2 hours of post-processing and registration operations. Therefore, these studies regarded point clouds as static assets, which need to be captured and preprocessed in advance so that can then serve for the subsequent real-time processes.

### 4.3.5 Multimodal methods

Multimodal methods refer to the combination of two or more methods mentioned for proximity perception. For example, Fang et al. [21] developed an active safety system for mobile crane lifting operations, utilizing rotation encoders, IMUs, and point clouds. Chen et al. [100] combined visual and point cloud techniques to align top-view 2D images with laser-scanned point clouds for real-time 3D bounding box estimation. Price et al. [79] developed a real-time crane monitoring system that integrates encoders, computer vision, and laser scanning to reconstruct a 3D workspace model of the crane environment, providing real-time spatial feedback to operators. Yong et al. [93] applied ultrasonic sensors to measure the distances of entities detected by a vision-based 2D object detection model.

### 4.4 Alarm issuing and receiving

The modes of alarms primarily include visual, auditory, and tactile alarms (also known as vibrotactile alarms), as well as combinations of these alarms.

Visual alarms can be shown to operators through monitoring displays in the cab [1,66], to workers through wearable devices (such as AR glasses) [88], or to all people through flashing light-emitting diodes (LEDs) [10,64,65]. For example, Kim et al. [88] visualized hazard information as colored arrows and displayed them to workers through AR glasses to indicate the direction and level of danger. The system developed by Kim et al. [65] triggers LED light strips attached to workers' helmets when it detects a dangerous proximity, providing visual warnings to both workers and drivers. Auditory alarms are typically issued by buzzers or speakers, which can be installed at fixed locations [2,26,103,104], on construction equipment [1], or on workers' wearable devices [4,31]. For example, Teizer et al. [4] developed a SmartHat that sends real-time alerts to the wearer through its accompanying buzzer. Martin et al. [31] used an electronic watch equipped with a piezoelectric buzzer to send vehicle intrusion alerts to workers in road construction zones.

However, the visual and auditory alarms sometimes have little effect since workers may be too focused on their work to notice these alarms. Additionally, the noisy construction environment worsens the situation, particularly in scenarios that require workers to wear earplugs or earmuffs. Vibrotactile alarms trigger workers' alertness by stimulating the human skin through vibration, effectively alleviating this problem [39,53,104]. For example, Sakhakami et al. [39] developed a wearable tactile system to indicate the direction and level of hazard by controlling the vibration combination and intensity of motors at different positions. Similarly, Huang et al. [53] designed various types of vibrations for different proximity situations. However, various combinations and multiple types of vibrations might make tactile alarms complex and confusing, increasing the learning cost for workers to adapt to the specific meanings of these alarms.

Different forms of alarms may produce varying warning effects, as workers react differently to them [2,103–105]. Investigating these differences can help determine the optimal strategy for issuing and receiving alarms. For example, Awolusi and Marks [103] examined the attenuation of two types of auditory alarms over distance and the response speed of workers to these alarms, concluding that workers responded faster to radio-based alarms than to pneumatic/microwave alarms. By quantifying habituation to auditory warnings using workers' behavioral and physiological data [26], Chae et al. [2] explored the effectiveness of three different auditory alarms in reducing operators' habituation to auditory warnings of construction equipment. Based on sufficient experiments, Yang and Roofigari Esfahan [104] found that workers' response times to visual and vibrotactile alarms were significantly shorter than to auditory alarms. They also compared the efficiency of different notification strategies in conveying directional information to users through tactile means, with corresponding metrics for evaluation. Combining multiple types of alarms might be the optimal strategy, as it can compensate for the limitations of applying only a single type of alarm, such as visual-auditory [1,10,66], auditory-tactile [10,41], visual-tactile [64], and visual-auditory-tactile [105] alarms. For example, Sabeti et al. [105] investigated how four different combinations of alarm modes affect workers' reaction time under the controlled outdoor workspaces and indoor virtual environments, concluding that auditory-tactile alarms elicit the fastest response, followed by visual-auditory-tactile alarms.

## 5. Discussion

This section primarily discusses the limitations and challenges of current proximity perception, as it is the foundation of proximity warning systems. Some potential future directions are subsequently mentioned to address these challenges.

### 5.1 Limitations and challenges

First of all, the equipment's roaring sounds provide very limited information about the proximity. Currently, sound-based methods can only identify the type of equipment and its approach or departure, but can not accurately estimate the specific distance between the equipment and the observer. Besides, sound-based methods are highly vulnerable to surrounding noise. A cruel fact is that construction sites often feature running a variety of equipment and activities, accompanied by various types of noise. The noisy construction environment significantly increases the difficulty of extracting sound features and recognizing sound events.

Most sensor-based methods rely on the propagation of waves (e.g., radio or ultrasonic) for distance perception. Therefore, sensor-based methods are sensitive to the material of obstacles. For example, radio wave-based sensors (e.g., RFID, BLE, UWB, etc.) are susceptible to metal obstacles. Unfortunately, almost all construction equipment comprises various metal components and casings. In addition, sensor-based systems require a large number of sensors. For example, Arslan et al. [8] installed approximately 200 BLE beacons in two buildings to obtain the spatiotemporal trajectories of workers. Ansaripour et al. [6] placed four UWB tags on a single vehicle to mitigate the interference from obstacles. The large number of sensors undoubtedly increases the cost and complexity of operating and maintaining sensor-based systems.

With the rapid development of deep learning, vision-based methods are gaining popularity due to their wide monitoring range and non-invasive features. Unlike sensor-based methods, the video/image records provided by vision-based methods can be used as evidence of accident accountability. However, current vision-based methods still have much room for improvement in distance estimation. For example, Kim et al. achieved distance estimation [94] and trajectory prediction [95] using monocular images, with distance errors of 0.9 and 0.95 meters, respectively. Wang et al. [96] achieved an average absolute error of 0.942 meters in distance estimation based on binocular stereo images. In addition, although some studies have been conducted on 3D object detection [3,89,100], end-to-end 3D object detection models have not yet been extensively explored, particularly monocular image-based models, which can offer fast inference speeds. The lack of relevant datasets with 3D annotations might be one of the main reasons contributing to this situation. Finally, the robustness of visual models under extreme conditions (e.g., severe weather or visual occlusion) and the issue of personal privacy of image data have also not been fully explored.

Although laser-scanned point clouds provide the most accurate distance estimation, their acquisition and processing are quite time-consuming [79,100] and cannot meet the requirements of real-time monitoring. Therefore, laser scanning has only been tentatively explored as an emerging technology in the early stage, but has not received in-depth study and application in this field, as discussed in Section 3.2. Additionally, the non-real-time nature of laser-scanned point clouds means they can only be regarded as static assets to support subsequent monitoring tasks. However, entities on construction sites are constantly changing as the construction progresses, including workers, construction vehicles, and buildings. Lastly, the cost of laser scanners is significantly higher than that of sensors and surveillance cameras, which is another important reason that limits their further application.

## 5.2 Future directions

Based on the above review and discussion, sensor-based, vision-based, and point cloud-based methods are expected to continue dominating proximity perception in the future. However, the emphasis may vary for different types of methods.

First, vision-based methods can focus on end-to-end distance estimation based on 3D object detection. End-to-end means integrating object localization and distance estimation into a single step, while 3D object detection directly outputs the 3D positions and dimensions of entities based on the captured images without relying on any pre-measured references or assumptions. It is encouraging that research has already begun on the development of relevant image datasets with 3D annotations [106], which may inspire further studies to thoroughly explore the application of end-to-end 3D object detection in this field. Therefore, end-to-end 3D object detection, especially the faster monocular 3D object detection, might be one of the promising future directions.

Second, although capturing point clouds is time-consuming, 3D construction scenes remain necessary and valuable because they provide the most comprehensive information about the built environment. Considering entities on construction sites are constantly changing, future research on point cloud-based methods should focus on accelerating data capturing, processing, and updating to achieve real-time 3D reconstruction and updates of dynamic construction scenes. In recent years, LiDAR technologies with real-time scanning capabilities and centimeter-level accuracy are expected to replace traditional 3D laser scanners. Given the massive scale of point clouds, research based on local point cloud operations and updates is also worth exploring in depth. Therefore, locking in the local point cloud area, recognizing the differences between point clouds in previous and subsequent frames, and updating the recognized local point cloud in the global 3D scene may be other potential research directions.

Finally, multimodal methods are also a feasible future direction, as they integrate the advantages of multiple methods to make the system robust to various extreme construction conditions. For example, the combination of surveillance cameras and sensors can reduce the number of sensors and the failure of pure vision-based methods caused by obstacle occlusion. Similarly, integrating surveillance cameras with LiDAR can effectively mitigate the failure problem of pure vision-based methods caused by blurred views resulting from extreme construction conditions. It can be imagined that the fusion of sensors, surveillance cameras, and LiDAR will achieve better performance. However, multimodal means more devices and larger data scales. Therefore, future multimodal research, while pursuing performance, also needs to focus on reducing the hardware and computing costs of the entire system.

## 6. Conclusion

Based on 97 articles published from 2010 to 2024, this study comprehensively investigates various efforts for PMW using the mixed-methods approach that combines quantitative bibliometric analysis and qualitative review. Quantitative bibliometric analysis reveals the technological evolution of PMW over time, as well as the five most influential authors in the field and the two research networks to which they are associated. The qualitative review categorizes and summarizes relevant works from four key processes of PMW system development: influencing factor study, hazard level definition and determination, proximity perception, and alarm issuance and reception. Finally, through the discussion of the limitations and challenges of current proximity perception, some future research directions are identified, such as end-to-end 3D object detection, real-time 3D reconstruction and updating for dynamic construction scenes, multimodal fusion, etc.

**Acknowledgment**

Acknowledgment

## Appendix A

Table A1. Detailed information regarding literature retrieval.

| Platform | Retrieval string | Results | Timestamp |
|---|---|---|---|
| Scopus | TITLE-ABS-KEY ( ( worker* OR architecture OR engineering OR construction OR aec OR infrastructure OR site OR building OR civil ) AND ( "str?ck" OR "str?ck-by" OR "collision detection" OR "collision monitoring" OR "collision prediction" OR "collision warning*" OR "collision alert*" OR "proximity detection" OR "proximity monitoring" OR "proximity prediction" OR "proximity estimation" OR "proximity warning*" OR "proximity alert*" OR "proximity perception" OR "proximity awareness" OR "distance detection" OR "distance monitoring" OR "distance prediction" OR "distance estimation" OR "distance perception" OR "distance awareness" OR "distance prediction" OR "proximity hazard*" OR "proximity sensing" OR "proximity analysis" OR "proximity alarm*" OR "collision hazard*" OR "collision avoidance" OR "collision accident*" OR "collision alarm*" OR "intrusion warning*" OR "intrusion alarm*" OR "intrusion alert*" OR anticollision OR anti-collision OR "collision risk*" OR "dangerous proximity" OR "hazardous proximity" OR "unsafe proximity" ) ) AND PUBYEAR > 2009 AND PUBYEAR < 2025 AND ( LIMIT-TO ( DOCTYPE,"ar" ) ) AND ( LIMIT-TO ( SUBJAREA,"ENGI" ) OR EXCLUDE ( SUBJAREA , "PHAR" ) OR EXCLUDE ( SUBJAREA , "ARTS" ) OR EXCLUDE ( SUBJAREA , "NEUR" ) OR EXCLUDE ( SUBJAREA , "ECON" ) OR EXCLUDE ( SUBJAREA , "PSYC" ) OR EXCLUDE ( SUBJAREA , "ENER" ) OR EXCLUDE ( SUBJAREA , "MEDI" ) OR EXCLUDE ( SUBJAREA , "BUSI" ) OR EXCLUDE ( SUBJAREA , "CENG" ) OR EXCLUDE ( SUBJAREA , "EART" ) OR EXCLUDE ( SUBJAREA , "BIOC" ) OR EXCLUDE ( SUBJAREA , "CHEM" ) OR EXCLUDE ( SUBJAREA , "ENVI" ) OR EXCLUDE ( SUBJAREA , "PHYS" ) OR EXCLUDE ( SUBJAREA , "MATE" ) OR EXCLUDE ( SUBJAREA , "MATH" ) OR EXCLUDE ( SUBJAREA , "SOCI" ) ) AND ( LIMIT-TO ( LANGUAGE,"English" ) ) | 1251 | Fri Jan 17 2025 18:30:59 GMT+0800 |
| WOS | TS=(worker* OR architecture OR engineering OR construction OR AEC OR infrastructure OR site OR building) AND TS=("str?ck" OR "str?ck-by" OR "collision detection" OR "collision monitoring" OR "collision prediction" OR "collision warning*" OR "collision alert*" OR "proximity detection" OR "proximity monitoring" OR "proximity prediction" OR "proximity estimation" OR "proximity warning*" OR "proximity alert*" OR "proximity perception" OR "proximity awareness" OR "distance detection" OR "distance monitoring" OR "distance prediction" OR "distance estimation" OR "distance perception" OR "distance awareness" OR "distance prediction" OR "proximity hazard*" OR "proximity sensing" OR "proximity analysis" OR "proximity alarm*" OR "collision hazard*" OR "collision avoidance" OR "collision accident*" OR "collision alarm*" OR "intrusion warning*" OR "intrusion alarm*" OR "intrusion alert*" OR anticollision OR anti-collision OR "collision risk*" OR "dangerous proximity" OR "hazardous proximity" OR "unsafe proximity") and Article (Document Types) and English (Languages) and Engineering (Research Areas) and Biophysics or Education Educational Research or Geochemistry Geophysics or History Philosophy Of Science or Nuclear Science Technology or Oceanography or Optics or Polymer Science or Public Administration or Radiology Nuclear Medicine Medical Imaging or Rehabilitation or Surgery or Astronomy Astrophysics or Mathematics or Chemistry or Social Sciences Other Topics or | 1204 | Fri Jan 17 2025 18:27:47 GMT+0800 |
| | Materials Science or Telecommunications or Geology or Environmental Sciences Ecology or Physics or Water Resources or Meteorology Atmospheric Sciences or Science Technology Other Topics or Energy Fuels or Neurosciences Neurology (Exclude – Research Areas) | | |

Table A2. Detailed information regarding keyword unification.

| **Final keyword** | **Relevant keywords** |
|---|---|
| 'accidents' | 'accidents', 'accidents, injuries, and fatalities', 'occupational accidents' |
| 'around-view monitoring' | 'around view monitor', 'around-view monitoring' |
| 'building information modeling (bim)' | 'bim', 'building information modeling', 'building information modeling (bim)' |
| 'blind-spot analysis' | 'blind spots', 'blind-spot analysis', 'equipment blind spots', 'vehicle blind spots' |
| 'bluetooth' | 'bluetooth', 'ble beacons', 'bluetooth beacon', 'bluetooth low energy (ble)', 'bluetooth technology' |
| 'buried utilities' | 'buried utilities', 'buried pipes' |
| 'collision hazards/accidents' | 'collision', 'collision accident', 'collision accidents', 'collision hazards', 'collision hazard analysis' |
| 'collision avoidance' | 'collision avoidance', 'collision prevention', 'collision avoidance of cranes' |
| 'collision warning' | 'collision warning', 'collision risk warning model', 'anticollision system' |
| 'construction equipment' | 'construction equipment', 'construction equipment and machines', 'construction vehicle', 'earthwork equipment', 'equipment', 'heavy construction equipment', 'heavy equipment', 'intelligent heavy equipment', 'mobile device' |
| 'construction safety' | 'safety', 'construction safety', 'construction safety management', 'construction site safety', 'safety and security', 'safety in construction site', 'safety management', 'safety risk management', 'work zone safety', 'work-zone safety', 'occupational health and safety', 'occupational safety', 'occupational safety and health design and planning auditing, education, and training' |
| 'construction sites' | 'construction site', 'construction sites' |
| 'mobile/tower cranes' | 'crane', 'cranes', 'crane operations', 'mobile crane', 'tower crane' |
| 'distance estimation' | 'distance estimation', 'distance assessment' |
| 'electrodermal activity (eda)' | 'electrodermal activity (eda)', ' eda' |
| 'hazard detection' | 'hazard detection', 'hazard perception' |
| 'information technologies' | 'information technologies', 'information technology' |
| 'internet of things (iot)' | 'internet of things (iot)', 'industrial iot' |
| 'intrusion alarms' | 'intrusion alarms', 'intrusion alert' |
| 'intrusion behaviors' | 'intrusions', 'intrusion behavior' |
| 'labor and personnel issues' | 'labor and personnel issues', 'labor' |
| 'laser scanning' | 'laser scanning', 'laser scanner' |
| 'object tracking' | 'object tracking', 'multiple object tracking' |
| 'object detection' | 'object detection', 'deep-learning-based object detection' |
| 'proximity warning systems' | 'personnel proximity warning system', 'proximity warning system', 'proximity warning systems', 'proximity detection and alert systems', 'proximity alerts' |
| 'pose estimation' | 'pose estimation', 'excavator three-dimensional (3d) pose estimation' |
| 'proactive safety' | 'pro-active safety', 'pro-active real-time safety' |
| 'proximity detection' | 'proximity detection and alert', 'proximity detection', 'proximity monitoring', 'proximity prediction', 'proximity sensing' |
| 'proximity hazards' | 'near miss and proximity hazard events', 'proximity and struck-by hazards' |
| 'radio frequency identification (rfid)' | 'rfid', 'radio frequency technology', 'zigbee rfid sensor network' |
| 'real-time location systems' | 'real-time location', 'real-time location system', 'real-time location system (rtls)', 'real-time location systems', 'real-time resource location tracking' |
| 'highway work zone' | 'roadway work zone', 'highway work zone' |
| 'sensing technologies' | 'sensing technology', 'sensing technologies', 'sensing' |
| 'sensors' | 'sensor', 'sensors' |
| 'struck-by hazards/risk/accidents' | 'struck by', 'struck-by', 'struck-by hazard', 'struck-by hazards', 'struck-by accident', 'stuck-by accident', 'struck-by risk' |
| 'ultra-wideband (uwb)' | 'ultra-wideband', 'ultra wideband', 'uwb' |
| 'unmanned aerial vehicle (uav)' | 'unmanned aerial vehicle', 'unmanned aerial vehicle (uav)' |
| 'vibrotactile warning' | 'vibrotactile warning', 'vibro-tactile alert' |
| 'virtual reality (vr)' | 'virtual reality', 'virtual reality (vr)' |
| 'workers' | 'workers', 'workers-on-foot', 'worker', 'construction worker', 'construction workers-on-foot' |